\documentclass[lettersize,journal]{IEEEtran}
\usepackage{amsmath,amsfonts}
\usepackage{algorithmic}
\usepackage{algorithm}
\usepackage{array}
\usepackage[caption=false,font=normalsize,labelfont=sf,textfont=sf]{subfig}
\usepackage{textcomp}
\usepackage{stfloats}
\usepackage{url}
\usepackage{verbatim}
\usepackage{graphicx}
\usepackage{cite}
\begin{document}

\title{Achieving Network Resilience through Graph Neural Network-enabled Deep Reinforcement Learning}

\author{Xuzeng Li, Tao Zhang,~\IEEEmembership{Member,~IEEE}, Jian Wang, Zhen Han, Jiqiang Liu,~\IEEEmembership{Senior Member,~IEEE}, 

Jiawen Kang,~\IEEEmembership{Senior Member,~IEEE}, Dusit Niyato,~\IEEEmembership{Fellow,~IEEE}, Abbas Jamalipour,~\IEEEmembership{Fellow,~IEEE}
\thanks{This work is supported in part by the Fundamental Research Funds for the Central Universities under Grant 2022JBZY020, in part by the National Natural Science Foundation of China (NSFC) under Grants 62402029, in part by the China Postdoctoral Science Foundation under Grant 2024T170047,
GZC20230223, 2024M750165. (Corresponding authors: Tao Zhang, Jian Wang)}

\thanks{Xuzeng Li, Tao Zhang, Jian Wang, Zhen Han and Jiqiang Liu are with the School of Cyberspace Science and Technology, Beijing Jiaotong University and also with the Beijing Key Laboratory of Security and Privacy in Intelligent Transportation, Beijing 100044, China (e-mail: 22110130@bjtu.edu.cn; taozh@bjtu.edu.cn; wangjian@bjtu.edu.cn; zhan@bjtu.edu.cn; jqliu@bjtu.edu.cn).}

\thanks{Jiawen Kang is with the Technology School of Automation, Guangdong University of Technology, China (e-mail: kavinkang@gdut.edu.cn).}

\thanks{Dusit Niyato is with the College of Computing and Data Science, Nanyang Technological University, Singapore (e-mail: dniyato@ntu.edu.sg).}

\thanks{Abbas Jamalipour is with the University of Sydney, Sydney NSW 2006, Australia (e-mail: a.jamalipour@ieee.org).}
}


\markboth{IEEE Network,~Vol.~XX, No.~XX, XX~2024}%
{Shell \MakeLowercase{\textit{et al.}}: A Sample Article Using IEEEtran.cls for IEEE Journals}


\maketitle

\begin{abstract}
Deep reinforcement learning (DRL) has been widely used in many important tasks of communication networks. In order to improve the perception ability of DRL on the network, some studies have combined graph neural networks (GNNs) with DRL, which use the GNNs to extract unstructured features of the network. However, as networks continue to evolve and become increasingly complex, existing GNN-DRL methods still face challenges in terms of scalability and robustness. Moreover, these methods are inadequate for addressing network security issues.
From the perspective of security and robustness, this paper explores the solution of combining GNNs with DRL to build a resilient network. This article starts with a brief tutorial of GNNs and DRL, and introduces their existing applications in networks. Furthermore, we introduce the network security methods that can be strengthened by GNN-DRL approaches.
Then, we designed a framework based on GNN-DRL to defend against attacks and enhance network resilience. Additionally, we conduct a case study using an encrypted traffic dataset collected
from real IoT environments, and the results demonstrated the effectiveness and superiority of our framework. 
Finally, we highlight key open challenges and opportunities for enhancing network resilience with GNN-DRL.

\end{abstract}


\section{Introduction}
As the number of mobile devices continues to grow, communication networks are expanding in scale, and user demands for the higher QoS are steadily rising. These trends present significant challenges for network management. Building a resilient network has long been a fundamental and important research direction in communication networks, with the primary goals of ensuring the quality of service (QoS), minimizing network overhead, and defending against network attacks.

To achieve better network management and enhance network resilience, deep reinforcement learning (DRL), as an automatic policy optimization method, has been widely adopted in various network control tasks, including network scheduling, computing power allocation, and routing optimization. Existing studies typically employ mathematical models, such as Markov decision processes, to represent different network control tasks and then automatically generate optimization policies through DRL. However, due to the large scale and heterogeneous architecture nature of networks, current DRL-based methods struggle to adapt to diverse network scenarios.

The powerful modeling and feature extraction capabilities of graph neural networks (GNNs) in handling unstructured data make them widely applicable across various domains. The inherent advantages of combining networks and GNNs are significant, as many relationships within a network can be effectively represented using graphs, including the physical topology, routing strategies, and user relationships. 
With the help of the relationship expression ability of graphs, the rich unstructured features can be extracted through GNNs to obtain accurate network status representation. 
Therefore, researchers have combined GNN and DRL, utilizing the network features extracted by GNNs, and then employing DRL to generate optimization policies for network optimization tasks \cite{huihongwei_2023_4543327}.
However, existing GNN-DRL research has limitations in that it  overlooks network security issues, such as the flood, distributed denial-of-service (DDoS) and brute force attacks within the network.

\begin{figure*}[]
	\centering
	\includegraphics[width=\linewidth]{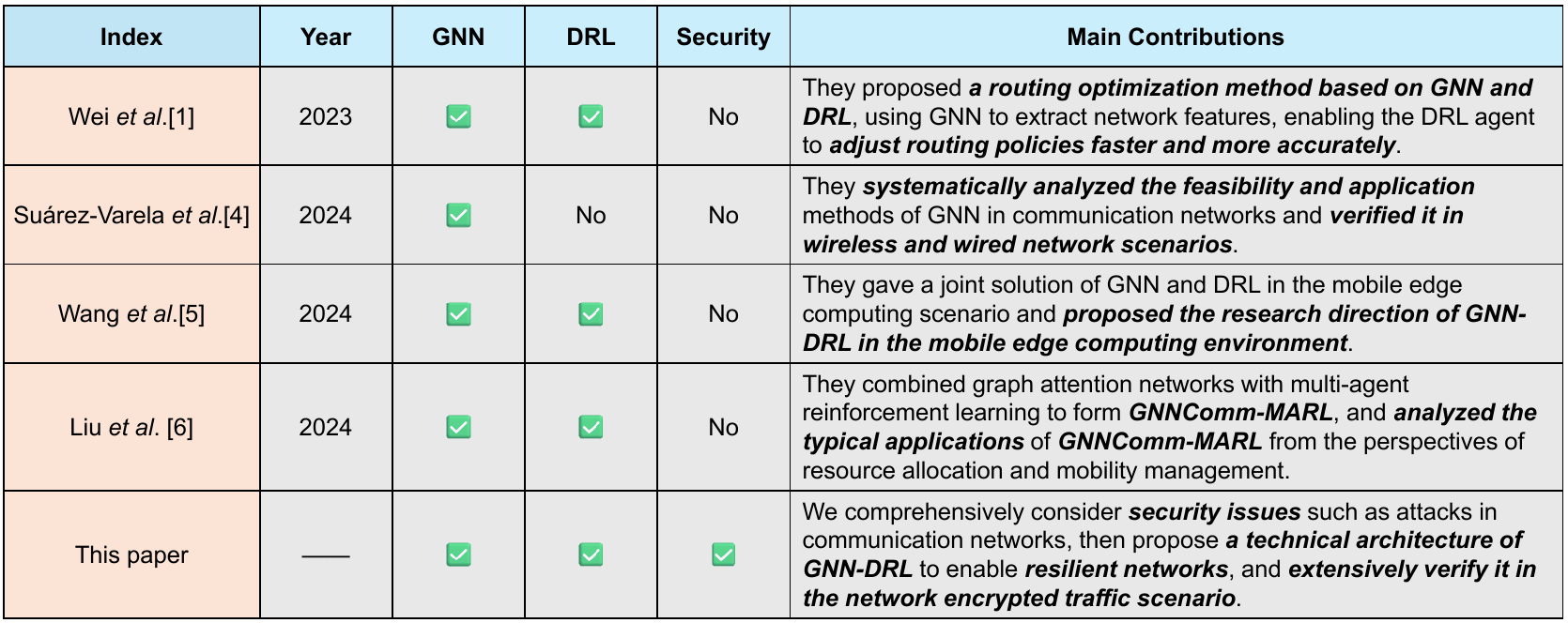}
	\caption{Comparison with existing researches on GNN and DRL in communication networks. We investigate the related works and select articles from the past two years that are closely related to GNN-DRL and communication networks for comparison. The index in the first row of the table indicate whether the research analyzed GNN, DRL, and network security. Most current studies focus on combining GNNs and DRL to improve the QoS, while research on using GNN-DRL to address network security and improve network resilience remains limited.}
	\label{fig:comwithexist}
\end{figure*}

Therefore, this paper aims to enhance network resilience by addressing security issues through GNN-DRL. GNNs demonstrate exceptional performance in network attack detection tasks, such as the detection of malicious encrypted traffic\cite{ZhangTFE}. Concurrently, DRL serves as an instrumental method for implementing essential network security solutions, including dynamic defense and moving target defense. In light of this, we construct the neural network part of DRL based on GNNs to form GNN-DRL. In GNN-DRL, GNNs are used to transform the input data into feature vector representations, which are then fed into a multilayer perceptron to generate the corresponding policies. We establish a framework that integrates network attack detection and defense mechanisms based on the proposed GNN-DRL. This framework improves the ability of networks to defend against attacks, which supports the establishment of a resilient network. The contributions of this paper can be summarized as follows:

\begin{itemize}
    \item We begin with a foundational tutorial on GNN-DRL and provide a brief overview of existing GNN-DRL methods in communication networks. Then, we analyze the technical aspects through which GNNs can enhance DRL in network tasks, and highlight the feasibility of leveraging GNN-DRL to create secure and resilient networks.

    \item We establish a framework based on GNN-DRL to enhance network resilience. Initially, we model network data using graphs to extract unstructured information, which allows us to obtain precise representations of environmental features. Leveraging this graph-based modeling, we employ GNNs for attacks detection. Next, we develop a security status assessment mechanism to evaluate the network. Ultimately, we utilize DRL to generate an optimal routing policy based on the security status of the network, enabling us to counteract attacks and improve network resilience.

    \item We conducted a case study on the proposed framework using an encrypted traffic dataset collected from the real  Internet of Things (IoT) environments, demonstrating that our framework can effectively withstand attacks such as flood and brute force attacks. The results show that our method is superior to the baseline.
\end{itemize}

Although there are existing studies on the application of GNN and DRL in networks (Fig.~\ref{fig:comwithexist}), the security issues of networks such as diverse and complex attacks still need to be further studied.

\section{GNN-DRL methods in networks}
\begin{figure*}[]
	\centering
	\includegraphics[width=0.98\linewidth]{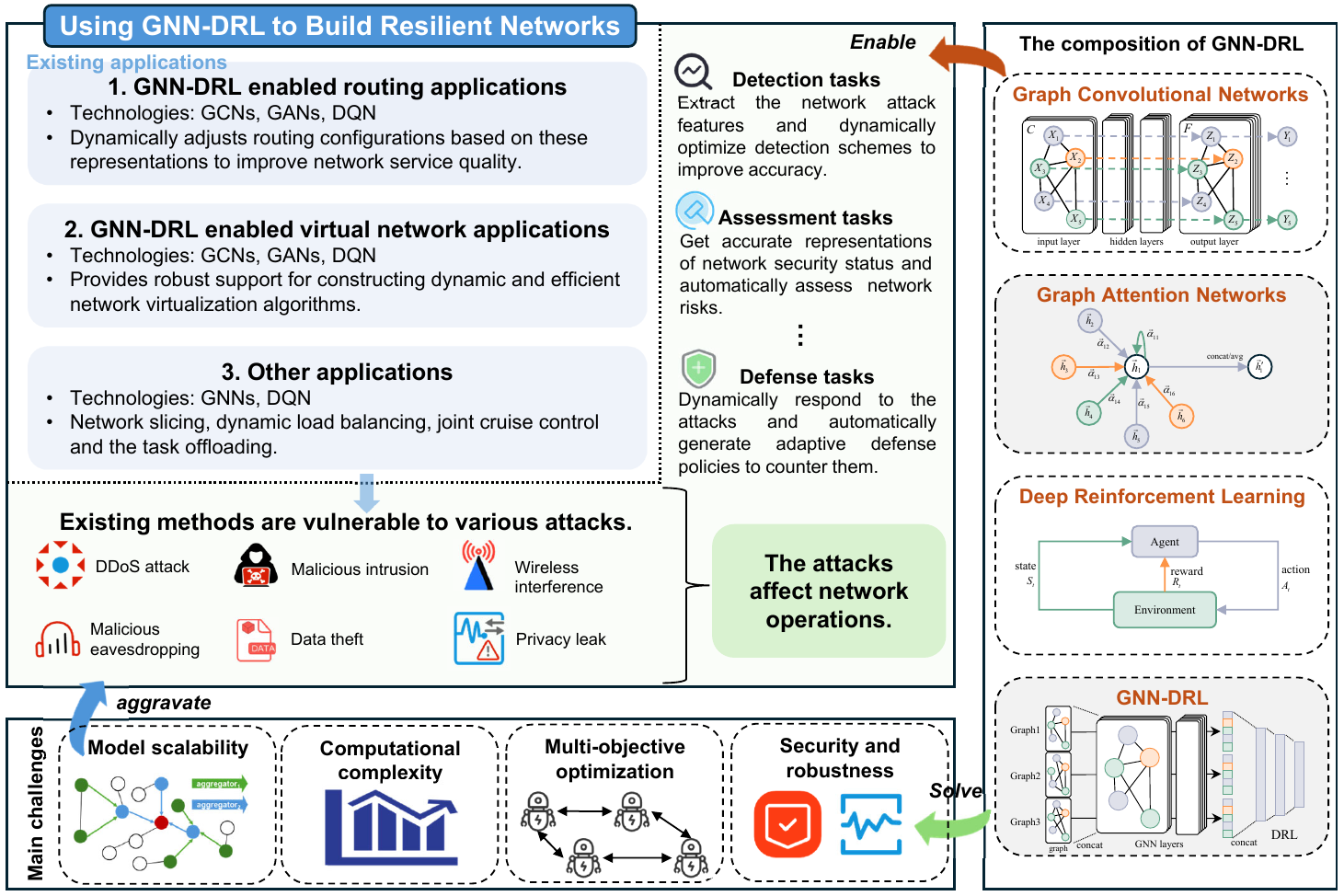}
	\caption{The schematic of building a resilient network based on GNN-DRL. Existing GNN-DRL methods for networks are vulnerable to diverse attacks. Moreover, designing GNN-DRL models for complex networks presents challenges such as model scalability, computational complexity, multi-objective optimization, and robustness. To address these issues, we explore GNN-DRL-based approaches aimed at enhancing network resilience by improving attack detection, risk assessment, and defense mechanisms.}
	\label{fig:schematic}
\end{figure*}

In this section, we first introduce representative algorithms in GNNs and DRL, along with various applications of GNN-DRL in networks. Next, we analyze the limitations of existing methods and propose potential development directions for GNN-DRL in building secure and resilient networks. Finally, we summarize the key challenges associated with utilizing GNN-DRL to improve the security and resilience of networks.  Fig. \ref{fig:schematic} illustrates the approaches of GNN-DRL for secure and resilient networks.

\begin{figure*}[]
	\centering
	\includegraphics[width=\linewidth]{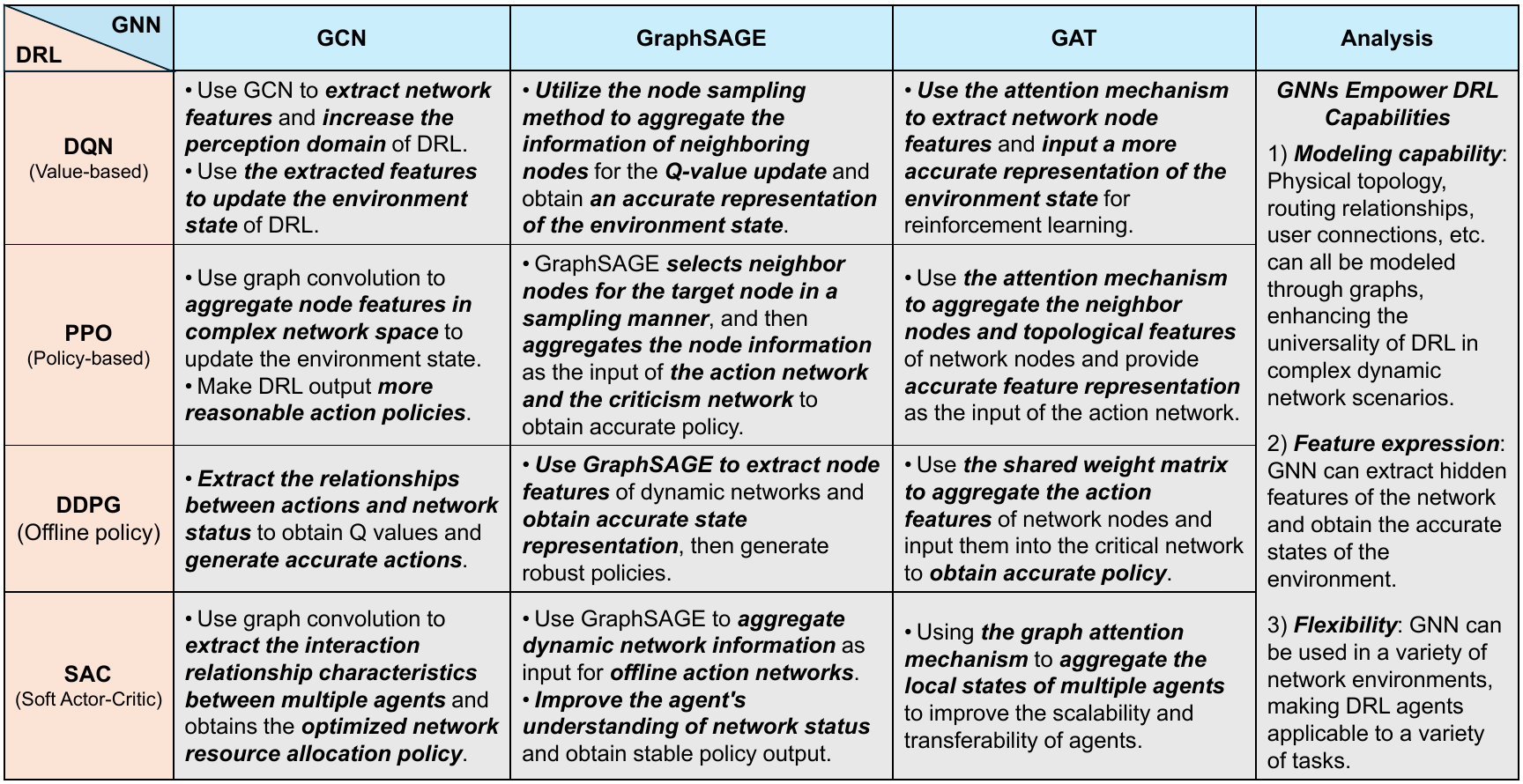}
	\caption{Analysis of different combinations of GNNs and DRL in communication networks. We selected representative models from GNNs (GCN, GraphSAGE and GAT), and combined them with different categories of DRL methods (DQN, PPO, DDPG and SAC) for analysis. Our findings suggest that GNNs can enhance the modeling and the feature extraction capability of DRL. Besides, GNN can improve the applicability of DRL across various network tasks.}
	\label{fig:summery}
\end{figure*}

\subsection{Fundamental GNNs and DRL methods}

\subsubsection{Fundamental DRL methods} DRL primarily addresses the challenge of determining the optimal policy and finds a wide range of applications in communication networks, including resource scheduling and routing optimization. The DQN algorithm is a notable representative of DRL. Unlike traditional reinforcement learning (RL) methods that establish a Q-value table, DQN employs a neural network to compute the Q-values for different status and actions, enabling it to make appropriate decisions. 
The policy-based methods aim to learn the target policy directly. Proximal Policy Optimization (PPO) is a policy gradient algorithm that improves upon previous calculations.
To address the issue of low sample efficiency in online policy algorithms such as PPO, offline policy algorithms like Deep Deterministic Policy Gradient (DDPG) have been proposed. DDPG constructs a deterministic policy and employs the gradient ascent method to maximize the Q-value, thus generating an optimized policy. Soft Actor-Critic (SAC) represents a more stable offline policy algorithm, offering effective sample learning while retaining the advantages of entropy maximization and enhanced stability.

Although DRL has demonstrated excellent capabilities in policy generation for communication networks, accurately representing network status and obtaining actionable feedback remain challenging due to the heterogeneous and dynamically changing nature of the network structure.

\subsubsection{Fundamental GNNs} GNNs are effective in processing unstructured data, and their powerful feature extraction capabilities and ability to represent heterogeneous graph data \cite{li2024permutation} have led to applications in various fields, including recommendation systems, social networks, and communication networks. In communication networks, graph neural networks are typically employed to uncover complex inherent information. Initially, the network is modeled as a graph, and accurate representations of network features are obtained through GNNs.
The Message Passing Neural Network (MPNN) is a flexible GNN specifically designed to operate on graph-structured data. MPNN utilizes a message-passing mechanism to aggregate and update node features by exchanging information between nodes and their neighbors. The key concept of the Graph Convolutional Network (GCN) is to generalize the convolution operation to graphs, enabling the model to aggregate information from neighboring nodes to learn node representations, which is a form of conductive learning \cite{josésuárez-varela_2023_4503840}.
GraphSAGE employs a sampling approach for target nodes to improve node feature aggregation patterns. Graph Attention Networks (GAT) employ an attention mechanism to assess the importance of neighboring nodes during information aggregation.

Researchers have recognized the network modeling capabilities, feature extraction abilities, and flexibility of GNNs in networks, leveraging these strengths in DRL to address its limitations. As shown in Fig. \ref{fig:summery}, we summarize the characteristics of various combinations of GNNs and DRL, and analyze how GNNs enhance the capabilities of DRL.

\subsubsection{Fundamental GNN-DRL methods} The fundamental architecture of the GNN-DRL method involves using GNNs to replace the neural network component in DRL. In the context of communication networks, GNNs typically process the environment state within the DRL model, extracting intrinsic features of the network to achieve an accurate representation of that state. Additionally, GNNs are employed to derive the reward values associated with actions or policies generated by the DRL, enabling the selection of optimal actions or policies.

\subsection{GNN-DRL methods in networks}
Networks including communication networks, wireless networks, mobile networks can be naturally represented as graphs from multiple views (e.g., topology, traffic and routing)\cite{josésuárez-varela_2023_4503840}. Applying GNN and DRL to enhance the network management has became more and more popular\cite{Wang10485450}\cite{Liu10638531}. Researchers use GNNs to obtain the precise representation of networks, and then combine it with DRL to generate optimized network configuration \cite{saimunikoti_2023_4528643}. The major applications of GNN-DRL in network can be divide into routing configuration, virtual network management and others.

\subsubsection{Routing configuration} The combination of GNNs and DRL has been widely applied in routing configuration. In typical scenarios, GNNs are utilized to derive network representations, after which DRL dynamically adjusts routing configurations based on these representations to enhance the QoS. Specifically, GNNs are employed to form representations of routers and the network topology as the state, and DRL is then used to optimize routing configurations\cite{huihongwei_2023_4543327}. GNNs can also model traffic behaviors, enabling methods to select efficient routing paths. Utilizing GNNs as the policy network, which provides a probability distribution over the actions of DRL, represents a novel approach\cite{saishreyasbhavanasi_2023_4543321}. In IoT, GNNs are employed to extract embedded information from the topology, while DRL agents identify abnormal connections\cite{yabinpeng_2022_4543318}.

\subsubsection{Virtual network management} Virtual network embedding is a fundamental aspect of network virtualization, and the integration of GNNs and DRL offers robust support for developing dynamic and efficient network virtualization algorithms. In general scenarios, GNNs are utilized to extract the status of both the physical and virtual networks. Subsequently, the actor and critic networks, or the policy network of a DRL agent, use this state as input to adjust the configurations of the virtual network, including the flow table, placement of function forwarding graphs, and other related elements \cite{yanghaoxie_2022_4529719}.

\subsubsection{Others} The combination of GNNs and DRL has many other applications, such as network slicing\cite{yanshao_2021_4530306}, offloading schedule\cite{kaili_2022_4543324}, and blockchain transaction optimizing\cite{jiahongcai_2023_4543325}.

Although existing GNN-DRL models can enhance the network, they often overlook the consideration of security. Current networks continually face various risks, including DDoS attack, malicious intrusions, data theft, and privacy breaches. The primary aim of this article is to investigate the synergistic integration of GNN and DRL to address these network threats and enhance overall network resilience.

\subsection{GNN-DRL for security on improving network resilience}

From the analysis in the previous section, we observe that existing GNN-DRL methods fall short in addressing events such as cyber attacks that significantly threaten network resilience. Consequently, we explore potential research directions for the GNN-DRL method aimed at enhancing network resilience with respect to network security.

\subsubsection{GNN-DRL helps network attack detection}
GNNs have been utilized in various network security detection tasks, including malicious traffic detection, malicious codes detection, and anomaly detection. By employing GNNs as a foundational attack detection method and integrating them with DRL, we can develop a novel architecture for attack detection that enables dynamic and accurate identification of network risks.
     
\subsubsection{GNN-DRL helps network security assessment}
GNN and DRL can be employed to dynamically assess the security status of the network and optimize its configuration. Specifically, GNN can generate a representation of the security status based on network characteristics. Then, DRL can refine the network configuration plan, allowing for real-time optimization of the configuration in response to the current network status.
     
\subsubsection{GNN-DRL helps network defense}
DRL is extensively utilized in network defense applications, including mobile target defense, security game theory, and security solutions for cyber-physical systems. When combined with GNNs, networks can accurately detect potential threats, after which DRL can formulate corresponding defense policies to enable dynamic and autonomous network protection.

\begin{figure*}[]
	\centering
	\includegraphics[width=0.99\linewidth]{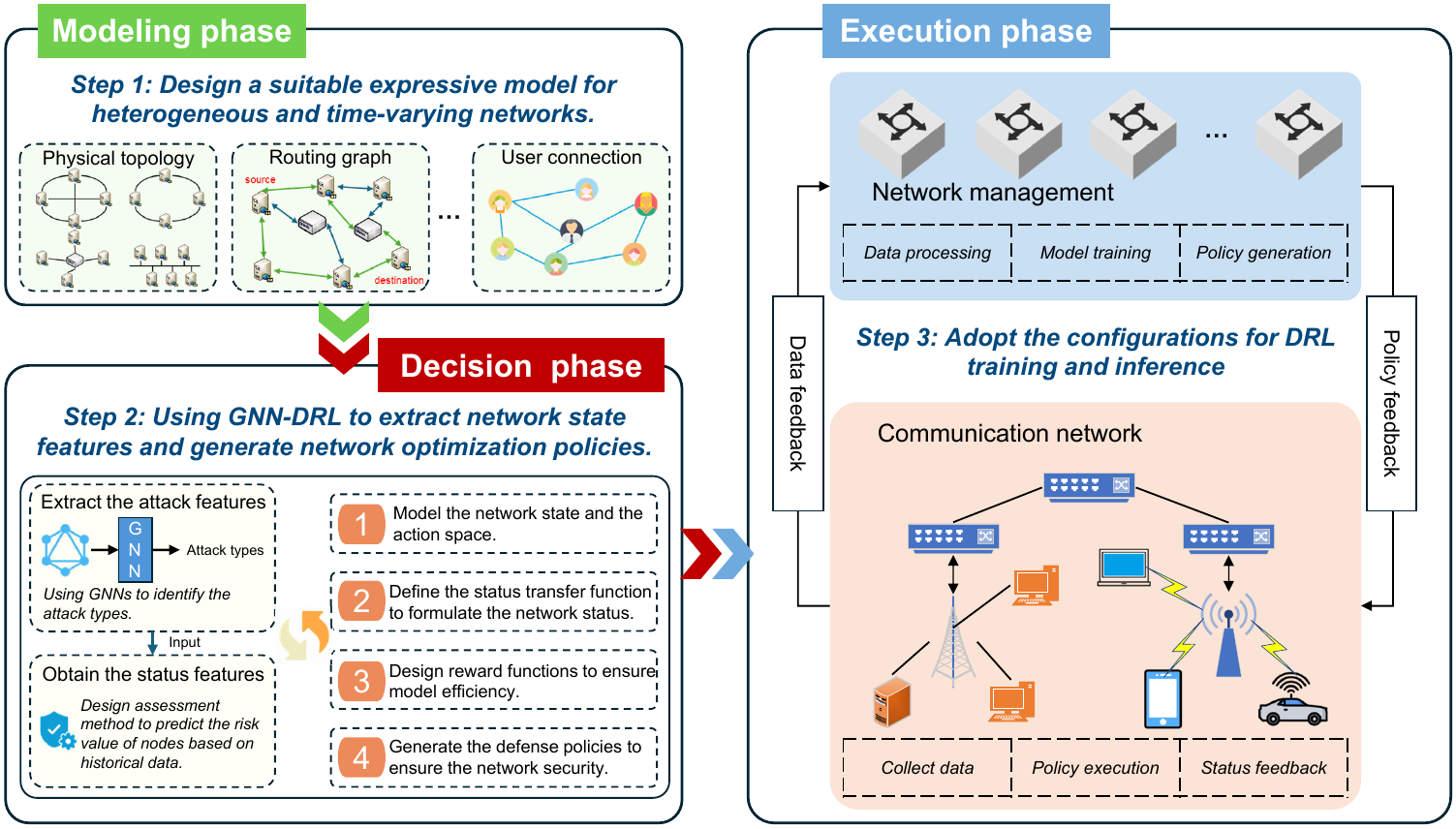}
	\caption{The framework of GNN-DRL for building secure and resilient networks. The framework comprises three phases: modeling, decision-making, and execution. During the modeling and decision-making phases, data processing is conducted, where network relationships are represented using graphs, and GNN-DRL methods are employed to generate optimized network configuration policies. In the decision-making phase, the network configuration is continuously refined based on the current network environment. This process involves generating configuration policies at the network management layer, which are subsequently applied to the network. The network then adjusts according to the optimized policies and provides feedback regarding the environment state.}
	\label{fig:framework}
\end{figure*}

\subsection{Main challenges of GNN-DRL methods for building a resilient network}

Using artificial intelligence algorithms to enhance network resilience involves addressing four significant challenges. Below, we analyze the open issues and opportunities associated with utilizing GNN-DRL to improve network resilience from four perspectives.

\subsubsection{Scalability of algorithms in large-scale networks}
Networks typically consist of numerous nodes and links, leading to a vast state space. GNN-DRL not only handles large-scale network graphs but also manages extensive state spaces, which imposes significant demands on computing resources, time complexity, and training efficiency. To tackle the scalability challenges of large-scale networks, the problem can be decomposed into multiple sub-problems. Additionally, enhancing the model's scalability can be achieved through techniques like distributed learning and parallel computing.

\subsubsection{Real-time and computational complexity challenges}
Decisions often need to be made quickly to satisfy the real-time requirements of the network. In large-scale networks, computing resources and response times can become bottlenecks during real-time inference. To mitigate these issues, the inference time of GNN-DRL can be reduced using techniques such as model compression, pruning, and knowledge distillation. Additionally, edge computing can be utilized to perform distributed inference closer to the data source, enhancing responsiveness and reducing latency.

\subsubsection{Multi-objective optimization and constraints}
Networks often involve multiple optimization objectives, such as latency, bandwidth, and energy consumption, which may present constraints or conflicts between different tasks. Effectively balancing these diverse requirements in multi-objective optimization problems poses a challenge for GNN-DRL applications. To address this, techniques from multi-objective DRL can be employed, or the weighted objective function method can be utilized to transform various objectives into a format that can be optimized simultaneously.

\subsubsection{Robustness challenges}
In an open communication network, network nodes and links may be subject to attacks or intrusions, leading to performance degradation of GNN-DRL systems. Furthermore, deep learning models themselves are vulnerable to adversarial attacks, adding further security requirements for their application in critical network tasks. Therefore, it is essential to design more robust GNN-DRL algorithms to mitigate the impact of adversarial attacks. Enhancing the security and robustness of GNN-DRL can be achieved through techniques such as adversarial training or the integration of defense mechanisms.

\subsection{Lessons learned}
The aforementioned cases illustrate the robust decision-making capabilities of GNN-DRL, motivating us to explore its application in resilient networks. Given that GNNs have already been effectively utilized in various network security contexts, we believe that GNN-DRL can be leveraged to address network attacks and enhance overall network resilience.

\section{The dynamic framework of GNN-DRL for building resilient networks}

Numerous studies have focused on detecting and defending against network attacks to enhance network resilience. However, the dynamic management of these attacks, along with integrated network optimization solutions for both detection and defense, remains an area that requires further exploration. In this section, we present the dynamic self-loop network resilience improvement architecture proposed in this paper. We outline how to utilize GNN-DRL to address security challenges in dynamically changing networks, focusing on three key stages: environment perception, policy generation, and policy execution.

\subsection{Technical designs}
\subsubsection{Graph-based large-scale complex network modeling}
The communication network can be modeled as an undirected graph $G=(V,E)$, where $V$ represents the set of nodes in the network, each node represents a communication device. Each node can add features to represent the properties of network nodes, such as communication capacity and computing performance. $E$ represents the set of edges, each edge represents a connection between network nodes. Each edge can be attached with a weight, representing the capacity, delay, or other characteristics.

\subsubsection{Network feature extraction based on GNN}
GNN takes graphs obtained through network modeling as input to perform feature extraction, so that it can capture the global and local structural information of the entire network. GNN aggregates the features of node $v$ with the features of its neighboring nodes $\mathcal{N}(v)$ through a multi-layer propagation mechanism, thereby obtaining the updated feature representation.

\subsubsection{The GNN-DRL method design}
The network state include node topology information, latency, traffic load and other features according to the task. Action refers to the operation that the agent can choose in each state. For example, the action can be to select a path or a link to transmit data in routing tasks. The action space can be designed as a discrete set of paths or a continuous amount of resource allocation according to the specific tasks. The reward can be related to network performance indicators such as latency, throughput, and energy consumption. It is critical to design a reasonable reward function, which needs to be able to motivate the agent to generate a better policy. Finally, the DRL algorithm is used to generate the policy 
$\pi(a|s)=\operatorname{argmax} Q\left(s_t, a\right)$, which determines the action taken in a specific state, where $Q\left(s_t, a\right)$ is the Q-value function based on the GNN extracted features.

\begin{figure*}[h]
	\centering
	\includegraphics[width=0.92\linewidth]{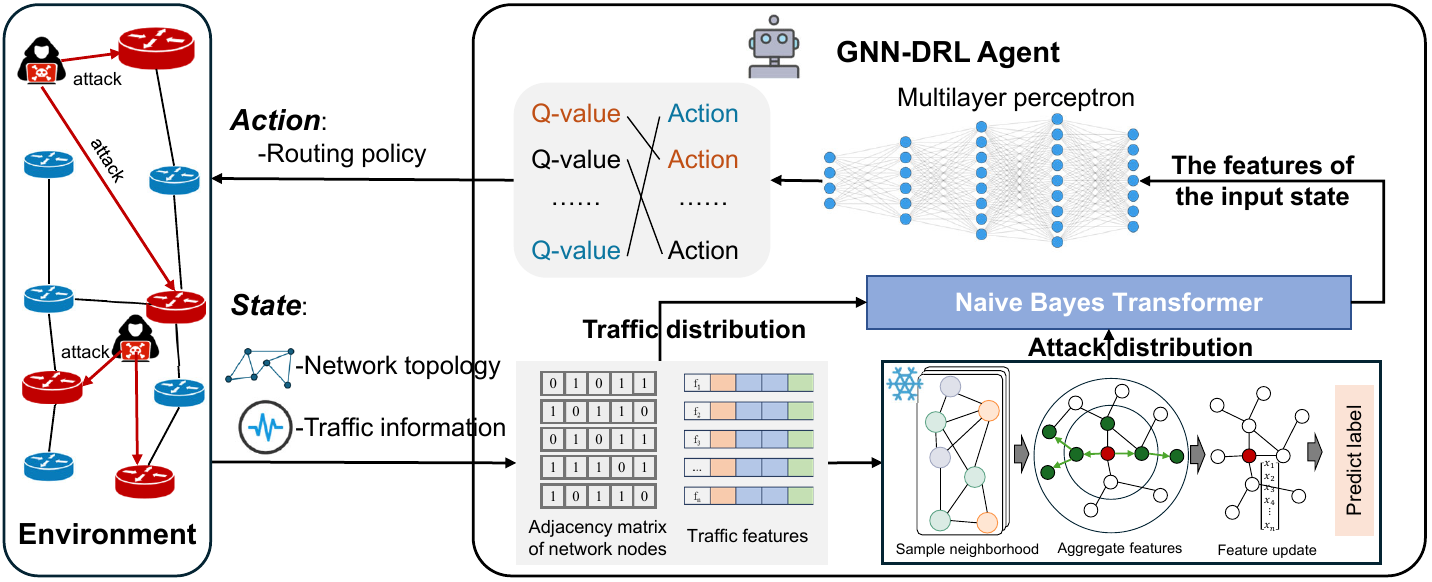}
	\caption{GNN-DRL for building a resilient and secure network. We utilize network topology and traffic as inputs, modeling encrypted traffic within the network using graphs. By applying GCN, we can effectively identify different types of malicious encrypted traffic. Additionally, we employ Bayesian methods to achieve a more accurate representation of the environmental status. Ultimately, we use DQN to generate routing policies that help avoid attacks.}
	\label{fig:case}
\end{figure*}

\subsection{Framework}
As shown in Fig. \ref{fig:framework}, we propose a general autonomous and dynamic network defense framework for achieving secure and resilient networks. The framework includes the modeling phase, the decision-making phase, and the execution phase.

\subsubsection{Modeling phase}

Heterogeneous network forms have always been an obstacle to network optimization, which not only limits the accuracy and efficiency of network feature extraction, but also undermines the versatility of network optimization models. Therefore, it is necessary to design an effective heterogeneous network modeling method to accurately represent original information of the network. In our architecture, we first use graphs to model the network.
\textbf{Step 1: Design a suitable expressive model for heterogeneous and time-varying networks.} We utilize graphs to represent various relationships within the networks, such as physical topology, routing modes, and user relationships. The focus is on addressing the heterogeneous nature of networks.

\subsubsection{Decision-making phase}

The decision-making stage is the core component of the network resilience improvement architecture, where appropriate optimization policies must be generated based on the network state. For instance, in the context of enhancing network security, this stage first requires an accurate assessment of the security status based on its current state. This involves obtaining precise representations of network features, which are then utilized to generate relevant network adjustment policies.
\textbf{Step 2: Using GNN-DRL to extract network state features and generate network optimization policies.} To achieve this, we utilize GNNs to enhance the environmental perception capabilities of DRL, enabling the generation of optimal policies. In the decision-making stage, we take the graph representing the network state, which was constructed in the modeling stage as the input of GNN. This allows us to extract characteristics of the network state, including attack features and structural attributes. 
After obtaining the network state characteristics, it is crucial to further efficiently and reasonably adjust the network configuration to maximize the network optimization benefits. In this framework, DRL algorithms such as DQN and PPO can be employed to generate the optimal network configuration based on resource constraints and elastic network requirements.

\subsubsection{Execution phase}
In the execution phase, deploying the GNN-DRL model at the network management layer. Network data, such as traffic, is collected and input into the model to generate optimization policies. These policies are then executed within the network environment, and the results are fed back into the system for further analysis and adjustment. \textbf{Step 3: Adopt the configurations for GNN-DRL training and inference.} Through continuous iterative optimization, the algorithm eventually converges, and the optimal configuration can be derived based on the network status.

\subsection{Lessons learned}
In actual scenarios, it is necessary to design a suitable graph-based network data modeling method to obtain the accurate representation of the environmental state, thus supporting the DRL agent in dynamically adjusting the network configuration that can effectively defend against attacks.

\section{Case Study: GNN-DRL avoids attacks to improve the network resilience}

In this section, we conduct a case study on encrypted traffic attacks to illustrate how GNN-DRL enhances network resilience. These attacks are complex and exhibit unstructured characteristics. Standard DL approaches often struggle to extract attack features and cannot adequately represent unstructured network structures. Consequently, it is essential to enhance the DL module within DRL to improve its ability to extract network features. Fig. \ref{fig:case} shows the architecture of the case study. We employ the GCN to construct the neural network within DQN. Then, we obtain an accurate representation of environmental state using Bayes transformer. Subsequently, we utilize DQN iteration to generate routing policies that can avoid attacks. To validate our approach, we reconstructed the network topology using a dataset collected from a real IoT that includes various types of attacks\cite{Dadkhah9851966}.

\subsection{Parameters setup}
The proposed GNN-DRL method extracts the characteristics of the network through GCN and generates the policies through DQN. The reward function considers two major factors. Firstly, the selected paths need to have low communication costs. Each node is associated with a communication cost $c$ (an integer in the range $[1, 200]$). Furthermore, we need to minimize the number of nodes that overlap between the paths $p_1$ and $p_2$, denoted as $O = |p_1 \cap p_2|$, to avoid excessive communication pressure on individual nodes. We incorporate a penalty term in the reward function where $O$ is multiplied by a penalty factor $\alpha$. Secondly, the generated policies must avoid attack nodes. By considering the attack traffic ratio, each node is assigned a security weight $w$ (an integer in the range $[1, 200]$). Therefore, the reward function \(R = -\sum_{n \in p_1} (w_n + c_n) - \sum_{m \in p_2} (w_m + c_m) - \alpha \cdot O\).

We design a GNN following TFE-GNN\cite{ZhangTFE}, using two-layer GCNs to extract encrypted traffic features and detect the attacks. The learning rate is 0.005, the batch size is 128, the cross entropy loss function and the Adam optimizer are used, and the weight decay is 0. We conducted experiments on two networks containing 20 nodes and 50 nodes, and the size of the action spaces are 464 and 39,038. The DQN is used to conduct the optimal path combination, with a learning rate of 0.001, a discount factor of 0.99, an initial value of the random action probability of 1, a random action probability attenuation coefficient of 0.995, and a final value of the random action probability of 0.01.

\subsection{Experimental scenario and the comparison method}
CICIoT2022\cite{Dadkhah9851966} is an encrypted traffic dataset collected from real IoT environments[1] and is widely used in various tasks. We reconstructed two network topologies from CICIoT2022, containing 20 nodes (25 edges) and 50 nodes (59 edges). The traffic conditions of each node are kept consistent with CICIoT2022. The inputs to GNN-DRL include historical traffic, network topology, and the communication costs of network nodes. We set two routing requirements from node 0 to 19 and from node 2 to 12. Our goal is to generate routing paths without attack nodes while minimizing the communication costs of these two paths. The advanced network management method MPNN-DQN\cite{ALMASAN2022184} based on GNN-DRL is used as our baseline.

\subsection{Performance evaluation of the routing policy generation}
\begin{figure*}[ht]
	\centering
	\includegraphics[width=0.92\linewidth]{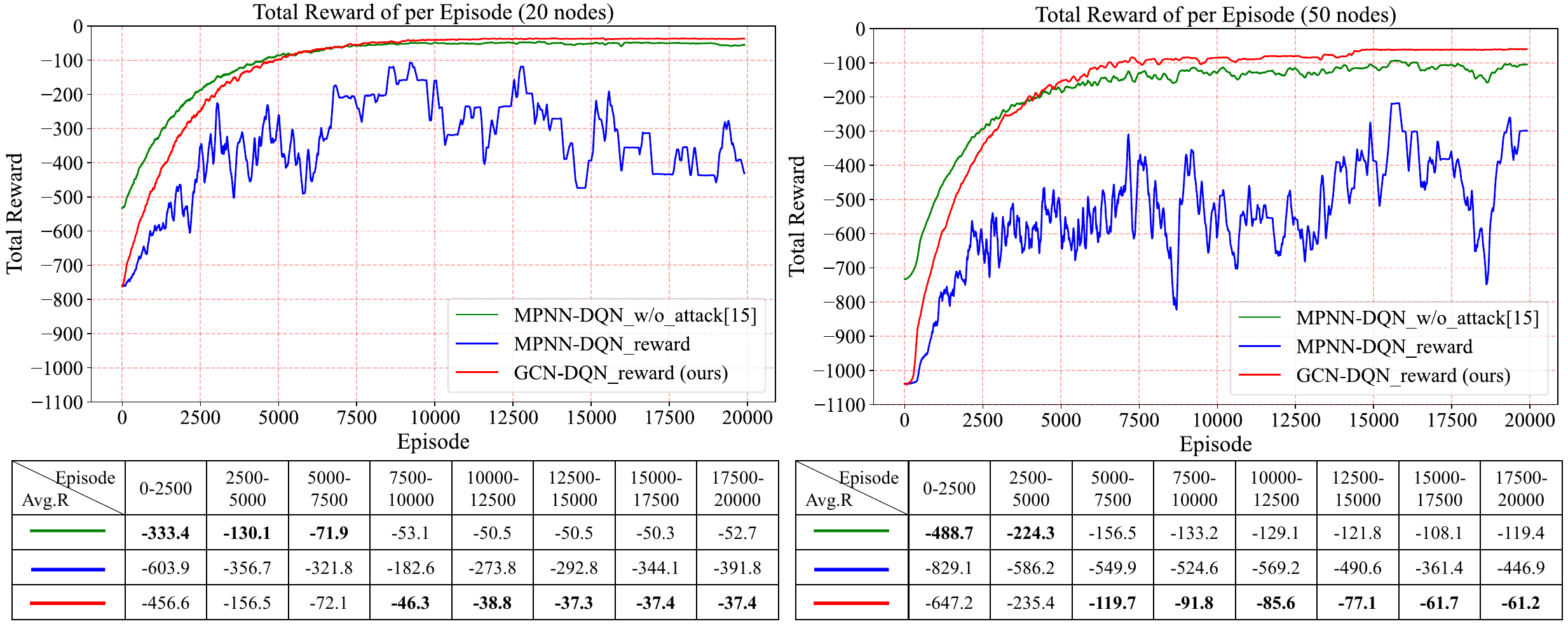}
	\caption{The comparison of reward values between our method with MPNN-DQN\cite{ALMASAN2022184} under different network scale (20 nodes and 50 nodes). All results were obtained by averaging over ten experiments. The curve is smoothed by calculating the local average using a sliding window (window size=100). The table below the graph records the average reward values of the algorithm within 2500 episode intervals, with the \textbf{bolded values} representing the optimal ones.}
	\label{fig:compare}
\end{figure*}

We conducted ten experiments, taking the average of values. The results are shown in Fig. \ref{fig:compare}. The green curve represents the performance of the MPNN-DQN\cite{ALMASAN2022184}, which considers only communication cost without accounting for attacks. The blue curve illustrates the performance of the MPNN-DQN in the presence of network attacks, while the red curve denotes the performance of our proposed method under the same attack conditions. Since MPNN-DQN cannot extract the attack features of the network, the reward of routing policies are lower (blue curve). Our method can converge quickly (red curve) and generate the optimal routing policy. The table in Fig. \ref{fig:compare} shows the optimal reward value of our method. Meanwhile, we evaluate the time consumption of GCN-DQN. The time consumption of GCN-DQN consists of the inference time and the time of policies generation. The GCN can process 1000 encrypted traffic packets within 0.49s. For networks with 20 nodes and 50 nodes, the time taken by the GCN-DQN is 0.031s per episode and 0.257s per episode, respectively.

The comparison of results shows that the MPNN-DQN method demonstrates oscillatory behavior, while our method achieves stable convergence to a higher reward, effectively reducing the impact of attacks.

\section{Future Directions}
Although some GNN-DRL methods have been employed to address network optimization problems, continuous developments and changes in network environments necessitate further optimization of GNN-DRL to better adapt to dynamic, large-scale, and diverse network scenarios.

\subsubsection{GNN-DRL for large-scale networks} The massive devices and complex relationships in large-scale networks will lead to huge graph for modeling, which will increase the difficulty of solving optimization problems. Multi-agent DRL sets up multiple agents in the environment to solve multi-objective problems. It can use multiple agents to cooperate to solve network optimization problems and split complex optimization problems into sub-problems to solve them step by step. Multi-agent DRL based on GNNs is a promising method for achieving large-scale network optimization.

\subsubsection{GNN-DRL in distributed network scenarios} Distributed networks represent a significant development trend. However, these networks face challenges including data leakage and sophisticated attacks including advanced persistent threats and zero-day vulnerabilities. Distributed learning approaches address these challenges by enabling computation without sharing original data. Combining GNN-DRL with distributed learning models offers promising future research directions. Integrating GNNs capable of detecting advanced attacks with DRL is also an important direction for further research. Meanwhile, exploring the integration of DRL with advanced network standards, such as the third generation partnership project, to promote its application in real-world networks is an important research direction.

\subsubsection{GNN-DRL under limited network resources} With the popularity of lightweight devices, there will be a large number of devices with limited or no computing power in the network. When the computing resources of network devices are limited, it is challenging to implement a real-time GNN-DRL algorithm. Using methods such as model pruning to optimize the GNN-DRL model structure and improve model efficiency is an issue that needs to be studied in depth.

\subsubsection{GNN-DRL in highly dynamic environments}
In rapidly changing network environments, where nodes and connections evolve rapidly, the GNN-DRL framework must not only quickly generate decisions, but also adapt to various network conditions. This presents challenges in terms of adaptability and computational time overhead.

\section{Conclusions}
In this paper, we summarize the existing GNN-DRL methods for communication networks and analyze the feasibility, significance and challenges of using GNN-DRL to improve network resilience. In addition, we emphasize the practicality and significance of using GNN to enhance DRL in the network. Then, we propose a resilient network framework based on GNN-DRL, which combines GNN and DRL to optimize network configuration and enhance network resilience. Experimental results on the dataset constructed from a real IoT environment demonstrate the advantages of our method. In addition, we point out the future directions of GNN-DRL in four different scenarios, including large-scale networks, distributed networks, limited network resources and highly dynamic environments.



\bibliographystyle{IEEEtran}
\bibliography{reference.bib}

\end{document}